\documentclass[11pt,a4paper]{article}
\usepackage{epsfig}
\usepackage{jheppub}

\newcommand{\be}{\begin{equation}}
\newcommand{\ee}{\end{equation}}
\newcommand{\bear}{\begin{eqnarray}}
\newcommand{\eear}{\end{eqnarray}}
\newcommand{\ba}{\begin{array}}
\newcommand{\ea}{\end{array}}

\title{Two Higgs Doublets from Fermion Condensation}

\author[a]{Gustavo Burdman}
\author[a]{Carlos E.~F.~Haluch}
\affiliation[a]{
Instituto de F\'{i}sica, Universidade de S\~{a}o Paulo,
\\ R. do Mat\~{a}o 187, S\~{a}o Paulo, SP 05508-900, Brazil }
\emailAdd{burdman@if.usp.br, haluch@fma.if.usp.br}

\abstract{
We consider the most generic situation in models where the electroweak
symmetry is broken by the condensation of a strongly
coupled fermion sector, such as for instance a fourth generation.  We study the scalar content resulting from
the condensation of both the up and the down type fermions,  corresponding  to a two-Higgs doublet model. We estimate the
scalar spectrum using the Nambu--Jona-Lasinio model, improved by the
renormalization group. We show that the scalar spectrum is generically
lighter than for the case with only one right-handed fermion condensing and that due
to a remnant Peccei-Quinn symmetry the lightest  state is the
pseudo-scalar, with masses ranging typically from $10~$GeV to
$120$~GeV.  We discuss the phenomenological consequences of this 
distinct spectrum. 
}


\begin{document}
 
\maketitle


\section{Introduction} \setcounter{equation}{0}
\label{intro}
The standard model (SM) gives  a very successful description of the gauge interactions of all
elementary particles known up to now~\cite{pdg}. On the other hand, the origin of the electroweak symmetry breaking (EWSB) sector, 
responsible for all mass scales, remains an open question. 
An alternative to the elementary scalar doublet introduced  in the SM,
is the possibility that a new interaction dynamically generates a new scale at low energies.
This paradigm, which is at work in the way the strong interactions
generate the hadronic scale, is used in technicolor theories˜\cite{tc1,strongreview}, where the new strong interaction resulting in
EWSB is confining. This results in a spectrum of ``technihadrons"
formed  by the strongly coupled fermions.
Alternatively, we could consider the possibility that the new strong interaction is spontaneously broken at around the scale where it becomes supercritical. 
In this case, the condensing fermions would not be confined. This raises the possibility of using SM fermions to break the electroweak symmetry through their
condensation~\cite{bhl}. However, the dynamical mass acquired by the
condensing fermions must be approximately $(500-700)$~GeV if we want
the scale at which condensation takes place to be close to
$O(1)$~TeV. Thus, we must introduce new fermions~\cite{fg1} in addition to the
new strong interaction. 
This   causes the condensates and leads to large 
dynamical masses for the condensing fermions, as well as to EWSB. 

In order to fully realize this scenario,  various ingredients must be
in place. In addition to the new fermions, a new strong interaction
spontaneously broken at the TeV scale and strongly coupled to them
must be present. 
Furthermore, higher-dimensional operators involving the massless fermions with the 
condensing ones must be present at low energies in order to feed masses and mixing to the SM fermions. 
A complete model was proposed in Ref.\cite{bd1} based on the SM with four generations in the bulk of a compact extra dimension with 
an AdS background. We will refer to this as the flavor AdS$_5$ model. 
This model, which minimally extends  bulk Randall-Sundrum models~\cite{rs1,gp} by adding a highly localized fourth generation in the bulk, constitutes
an existence proof as well as a source of detailed predictions~\cite{bdfgq,bdfglep}. 

In Ref.˜\cite{bd1} it was assumed for simplicity that 
only one of the fourth-generation quarks condenses. This leads to a composite scalar sector that corresponds to that of one scalar doublet, just as in the SM. 
The Higgs mass in that case was at the edge of perturbative unitarity. 
However, it is more natural to assume that both up and down-type
fermions  form condensates since the new strong interaction should largely respect isospin
symmetry in order to avoid $T$ parameter constraints. If both up and
down-type fermions condense, the resulting composite scalar spectrum
corresponds to that of a two-Higgs doublet model~\cite{fgcon}. 
In this paper we will study this possibility, particularly the low
energy scalar theory resulting from double condensation. 
Although we are taking inspiration from the flavor AdS$_5$ model to generate the necessary interaction among fourth-generation quarks, our 
results are valid any ultra-violet (UV) completion of the four-fermion interactions giving rise to the condensation. 
Thus, the resulting low energy scalar spectrum and phenomenology
should correspond to a large class of theories, independently of
whether the condensing fermions are fourth generation quarks or just
generic new
fermions strongly coupled to a spontaneously broken gauge interaction. 

The rest of the paper is organized as follows: in Section~\ref{model} we specify the new fermion interactions introduced 
the couplings to  massive gauge bosons. Several properties of the resulting scalar spectrum at energies below the gauge boson mass
are specified already here. In Section~\ref{scalspec} we compute the low energy scalar spectrum. We do this by solving the renormalization group
equations (RGE) for the scalar self-couplings and the fourth-generation Yukawa couplings. 
In Section~\ref{ewpc} we study the electroweak precision constraints in a model with four fermion generations and the scalar spectrum obtained in 
Section~\ref{scalspec}. The constraints from flavor physics and direct searches are discussed in Section~\ref{pheno}, whereas we conclude in 
Section~\ref{conc}.

\section{A Two-Higgs Doublet from Fermion Condensation}
\label{model}
We consider a scenario with a set of new fermions coupled to a new
strong interaction. This is mediated by massive gauge bosons, possibly
remnant from the spontaneous breaking of a gauge symmetry at the TeV
scale. For instance, in the flavor AdS$_5$ models mentioned in the previous section, the
strongest interaction among the fourth-generation quarks is provided
by the KK gluons and the fermions are the zero modes of
fourth-generation quarks. 
More generally, we can consider that there is a  new interaction spontaneously broken above the TeV scale, which 
will be responsible for fermion condensation. Thus, the minimal
fermion content we consider is given by
\be
Q^a=\left(\ba{c} U^a\\D^a\ea\right)_L, \quad
U^a, \, D^a~,
\label{matcont}
\ee
where the index $a$ refers to the new interaction felt by the
$SU(2)_L$ doublet $Q$ as well as by the singlets $U$ and $D$. 
The fermions in (\ref{matcont}) are the ones that will
condense. Additional matter content might be necessary in order to
cancel anomalies. For instance, if the fermions in (\ref{matcont}) are
fourth-generation quarks, leptons must be added in order to get an
anomaly free theory. Just as in this case, we assume that in general
the additional fermions needed to cancel anomalies will not
participate in condensation.
The new interaction  could  also couple to the lighter quarks,
although more weakly. Here we will not consider these interactions for
simplicity. 
We further assume that the new strong interaction is spontaneously
broken and that massive gauge bosons are integrated out to lead to
four-fermion interactions of the form
\bear
{\cal L}_{\rm eff} &=& \frac{g_L g_u}{M_G^2}\, \bar Q\gamma_\mu
T^a Q\,\bar U\gamma^\mu T^a U\nonumber\\
&+&   \frac{g_L g_d}{M_G^2} \,\bar Q\gamma_\mu
T^a Q \,\bar D \gamma^\mu T^aD~.
\label{fgffermion}
\eear 
where $T^a$ are the symmetry generators, and $g_L$, $g_u$ and $g_d$ are the gauge couplings of $Q$, $U$ and
$D$, respectively.
For instance, in the case where the new fermions are fourth-generation
quarks, the $T^a$ are  the $SU(3)_c$ generators, and the massive gauge
boson is a color-octect, just as in the AdS$_5$ flavor model of Ref.~\cite{bd1}.
Here we will consider a generic case, since the nature of the new
interaction as well as the identity of the condensing fermions is not
crucial in determining the low-lying scalar spectrum, as we will see below. 

The four-fermion interactions of interest leading to EWSB are 
\be
{\cal L}_{\rm eff.} = G_U \bar Q U\bar U Q + G_D \bar  Q D
\bar D Q
\label{ffermion4}
\ee
where the Nambu--Jona-Lasinio (NJL)  couplings are just 
\be
G_U\equiv \frac{g_L g_u}{M_G^2}, \qquad\quad G_D\equiv \frac{g_L
  g_d}{M_G^2}~.
\ee
In order to generate non-trivial values for the quark condensates $\langle\bar Q U\rangle$
and $\langle\bar Q D\rangle$, the couplings $G_U$ and/or $G_D$ must
be above some critical value.  
In Ref.~\cite{bd1} only one of the couplings ($G_U$) was considered
supercritical, leading to a composite scalar spectrum consisting of
one Higgs doublet. Here we consider a more general case, with both
couplings supercritical. As mentioned earlier, this results in a two-Higgs doublet
composite scalar spectrum, with very distinct masses. This scenario
will turn out to be quite different from the one considered in
\cite{bd1}, even if we just consider the neutral scalar masses. 

The four-fermion interactions can be bosonized by introducing the auxiliary fields $\tilde\Phi_U$ and $\Phi_D$, which are
$SU(2)_L$ doublets with hypercharges $h_U=-1/2$, and $h_D=1/2$.
Then the  NJL interactions of (\ref{ffermion4}) can be shown to be equivalent to 
\bear
{\cal L}_{\rm eff.} = Y_U (\bar Q \tilde\Phi_U U+ {\rm h.c.} )
+ Y_D (\bar Q \Phi_D D+ {\rm h.c.}  ) -M_G^2 \Phi^\dagger_U
\Phi_U  - M_G^2\Phi^\dagger_D\Phi_D~,
\label{lboson1} 
\eear
with the appropriate identifications:
\be
 Y_U^2 = g_L g_u, \qquad Y_D^2 = g_Lg_d~.
\ee

The auxiliary fields in (\ref{lboson1}) defined at the scale $M_G$
 develop kinetic terms and self-interactions  below that scale
\be
{\cal L}_{\rm kin.} = Z_{\Phi_U} (D_\mu\Phi_U)^\dagger D^\mu\Phi_U + Z_{\Phi_D} (D_\mu\Phi_D)^\dagger D^\mu\Phi_D
~,\ee
where the field renormalizations obey $Z_{\Phi_i}(\mu)\to 0$ as
$\mu\to M_G$. This leads to the renormalization of the parameters of
the potential. 
For instance, at leading order, the Yukawa couplings renormalize according to 
\bear
Y_U \to \frac{Y_U}{\sqrt{Z_{\Phi_U}}}, &&\quad Y_D \to
  \frac{Y_D}{\sqrt{Z_{\Phi_D}}}, \quad \nonumber
\label{rencoups} 
\eear
The mass terms, on the other hand, will get corrections that are quadratic in the cutoff $M_G$, leading to new renormalized masses at the scale $\mu<M_G$:
\bear
\mu_U^2 &=& M_G^2 - \frac{g_L g_u N_g}{8\pi^2}\left(M_G^2-\mu^2\right)\label{mu2}\\
\mu_D^2 &=& M_G^2 - \frac{g_L g_d N_g}{8\pi^2}\left(M_G^2-\mu^2\right)\label{md2}~,
\eear
where $N_g$ is the size of the fermion representation in the broken
group. For instance, if the new fermions are fourth-generation quarks
$N_g=N_c$, the number of colors. Finally, there will be scalar
self-interactions generated by fermion loops, which also get
renormalized by $Z_{\Phi_{U,D}}$. We will consider them in detail in
the next section, where we study the scalar potential.
 
As we will see below, the minimization condition on the scalar potential will almost certainly require that, for $\mu\ll M_G$, 
both $m_U^2<0 $ and $m_D^2<0$ be satisfied in order for $\Phi_U$ and $\Phi_D$ to get a VEV each, $v_U$ and $v_D$ respectively. 
In terms of the underlying gauge theory,  this translates into a criticality condition for the gauge couplings
\be
g_L  g_u > \frac{8\pi^2}{N_g}, \qquad  g_L  g_d > \frac{8\pi^2}{N_g}, 
\label{supercrit}
\ee
which  coincides with the conditions to make the NJL couplings in (\ref{ffermion4}) supercritical to form both fermion condensates, 
$\langle\bar Q  U\rangle$ and $\langle\bar Q D\rangle$. 

We notice that the tree-level auxiliary
lagrangian (\ref{lboson1})  {\em lacks} a term mixing $\Phi_U$ with $\Phi_D$.  This
means that the theory is invariant under the Peccei-Quinn symmetry 
\bear
Q \to e^{-i\theta} Q &&\quad U  \to e^{i\theta} U \qquad D\to
e^{i\theta} D \nonumber\\
\Phi_U\to e^{2i\theta} \Phi_U && \Phi_D\to e^{-2i\theta}\Phi_D~,
\label{pqsym} 
\eear
with $\theta$ an arbitrary phase. The mixing term in the bosonized
theory has the form
\be
{\cal L}_{\rm mix} = \mu_{UD}^2 (\Phi^\dagger_U\Phi_D  + {\rm h.c.} )
\label{lmix1}
\ee
and, although is not generated by fermion loops perturbatively, it
will be generated by instanton effects. As we will see in the next section, in the
$\mu_{UD}=0$ limit the pseudo-scalar state becomes massless. Instanton 
effects associated to the new strong interaction will lift this
``axion''  mass, but it will remain the lightest
state in the spectrum. 
Going back to fermion degrees of freedom, the term leading to the mixing written purely
in terms of fermion fileds is 
\be
{\cal L}_{\rm mix} = G_{UD} (\bar Q D \bar U^c\tilde Q + {\rm
  h.c.} ) ~,
\label{lmix2}
\ee
with 
\be
\tilde Q \equiv  - i\sigma_2 Q~.
\ee
This term is generated by instantons and in fact 
 is just the 't'Hooft flavor determinant \cite{tooft}
\bear
{\cal L}_{inst.} &= &\frac{\kappa}{M_G^2} {\bf det}\left[ \bar Q_{L} Q_{R} 
\right]\nonumber\\
& =&  \frac{\kappa}{M_G^2} \left[\bar (D_{L}D_{R})(\bar
  U_{L}U_{R}) - (\bar U_{L}D_{R})(\bar D_{L}U_{R})  +{\rm h.c.}\right]
\label{tooftdet}
\eear
where $\kappa$ is an $O(1)$  dimensionless coefficient  depending on details of
the instanton calculation. Thus, the mixing parameter in (\ref{lmix2})
$G_{UD} = \kappa/M^2_G$
is determined by instanton effects resulting from the new strong
interaction with a suppression energy scale $M_G$. This is similar to
the instanton effects found in Topcolor theories in
Ref.~\cite{topc1}. 
In our calculations we will take $\kappa$ to be a free parameter varying in the region $\kappa\simeq (0.1-1)$ .

\section{The Scalar Mass Spectrum}
\label{scalspec}

With the inclusion of the instanton-generated mixing term the most general potential that is generated by the fermion loops is given by
\bear
V(\Phi_U, \Phi_D)& =& \mu^2_U|\Phi_U|^2 + \mu^2_D|\Phi_D|^2 +
\mu^2_{UD}(\Phi_U^\dagger\Phi_D + {\rm h.c.} )\nonumber\\ 
&& + \frac{\lambda_1}{2} |\Phi_U|^4 + \frac{\lambda_2}{2} |\Phi_D|^4 
+ \lambda_3 |\Phi_U|^2 |\Phi_D|^2 + \lambda_4 |\Phi_U^\dagger\Phi_D|^2
\label{pot1}~,
\eear
where the scalar self-couplings $\lambda_i$, with $i=(1-4)$ are
generated by fermion loops. The potential in (\ref{pot1}) is not the most general 2HDM potential~\cite{thdmgen}. 
In general there would be other self-coupling terms allowed by the symmetries. However, these are not generated 
at one-loop level by the Yukawa interactions in (\ref{lboson1}).

The propagation of the scalars at low energies renormalizes the Yukawa
couplings, leading to a renormalization group evolution given by
\bear
\frac{dY_U}{dt} &=& \frac{1}{16\pi^2}\left[ Y_U^3 N_c + \frac{3}{2}
  Y_U^3 +\frac{1}{2}Y_D^2Y_U - C_U(t)Y_U\right]\label{yurge}\\
\frac{dY_D}{dt} &=& \frac{1}{16\pi^2}\left[ Y_D^3 N_c + \frac{3}{2}
  Y_D^3 +\frac{1}{2}Y_U^2Y_D - C_D(t)Y_D\right]\label{ydrge}~,
\eear
where $t=\ln(\mu)$, and the functions $C_U(t)$and $C_D(t)$
give the contributions from the SM gauge sector. They are given by
\bear
C_U(t)& =& 8g_s^2(t) + \frac{9}{4}g^2 +\frac{17}{12}g^{'2}\nonumber\\
C_D(t)& =& 8g_s^2(t) + \frac{9}{4}g^2 +\frac{5}{12}g^{'2} 
\label{gaugeconts}~,
\eear
with $g_s,g$ and $g'$ the $SU(3)_c$, $SU(2)_L$ and $U(1)_Y$ gauge
couplings respectively. We will assume that the new strong
interactions in (\ref{fgffermion}) do not break isospin symmetry. Thus, 
the only isospin breaking is introduced by the $U(1)_Y$ hypercharge
interaction in (\ref{gaugeconts}).  We can then safely approximate
\be
Y_U\simeq Y_D\label{ueqd}
\ee
which results in a nearly degenerate fourth-generation quark doublet. 
With this approximation, we can easily solve the RGEs for the Yukawa
couplings, resulting in
\be
Y_U(\mu) = Y_D(\mu)
\sqrt{\frac{C(\mu)}{5\left[1-\left(\frac{\mu}{\Lambda}\right)^{2 C^2(\mu)/16\pi^2}\right]}} 
\label{yukawavsmu}~,
\ee  
where we defined $C_U(\mu)\simeq C_D(\mu)=C(\mu)$ by neglecting the small
hypercharge contributions. For instance, for $\mu=M_Z$, $C_U$ and $C_D$
differ by less than $1$\%.

\begin{figure}
\begin{center}
\includegraphics[scale=1.1]{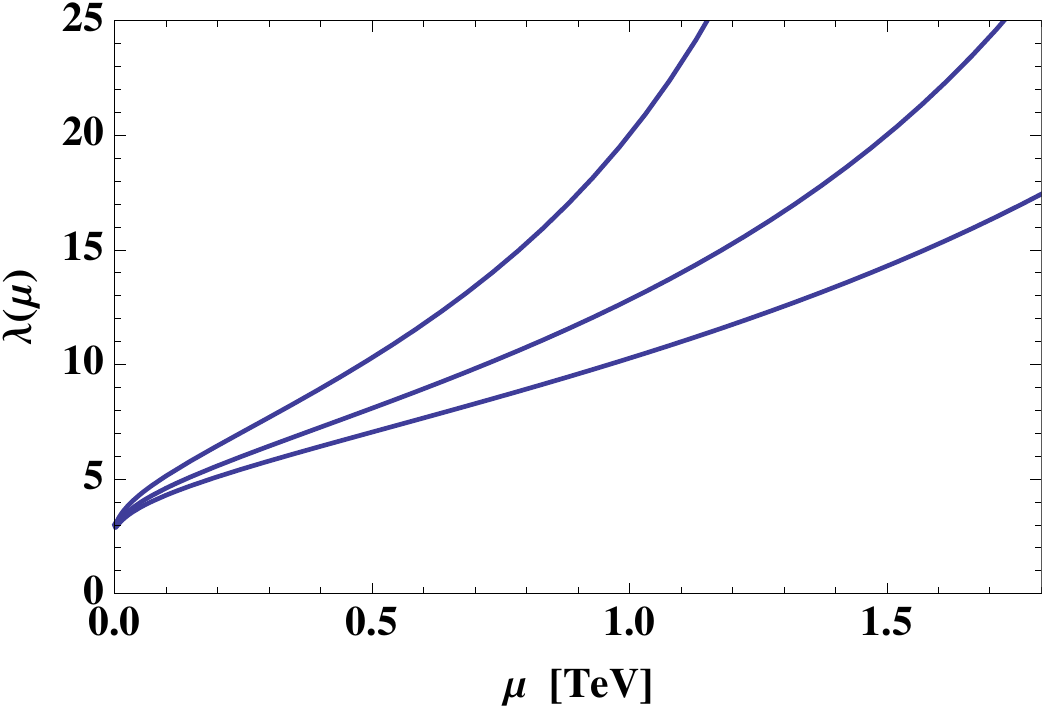}
\caption{Scalar self-coupling $\lambda_1$ as a function of energy, for
  a various choices of the cutoff scale $M_G$. From the top $M_G=2, 3,
  4$~TeV.} 
\end{center}
\label{lambda1}
\end{figure}

Next, we consider the RGEs for the renormalized scalar
self-couplings. These are given by 
\bear
\frac{d\lambda_1}{dt} &=& \frac{1}{16\pi^2} \left[ 12\lambda_1^2 +
  4\lambda_3^2  + 4\lambda_3\lambda_4
  +2\lambda_4^2-3\lambda_1(3g^2+g'^{2})+ \frac{3}{2}g^4 +
  \frac{3}{4}(g^2+g'^{2})^2 \right.\nonumber\\
&& \left. +12\lambda_1 Y_U^2 -12 Y_U^4\right] \label{rgel1}\\
\frac{d\lambda_2}{dt} &=& \frac{1}{16\pi^2} \left[ 12\lambda_2^2 +
  4\lambda_3^2  + 4\lambda_3\lambda_4
  +2\lambda_4^2-3\lambda_2(3g^2+g'^{2})+ \frac{3}{2}g^4 +
  \frac{3}{4}(g^2+g'^{2})^2 \right.\nonumber\\
&& \left. +12\lambda_2 Y_D^2 -12 Y_D^4\right] \label{rgel2}\\
\frac{d\lambda_3}{dt} &=& \frac{1}{16\pi^2} \left[
  (\lambda_1+\lambda_2)(6\lambda_3+2\lambda4)  +
  4\lambda_3^2  + 2\lambda_4^2
  -3\lambda_3(3g^2+g'^{2})+ \frac{9}{4}g^4 +
  \frac{3}{4}g'^{4} -\frac{3}{2}g^2g'^2  \right.\nonumber\\
&& \left.  +6\lambda_3(Y_U^2+Y_D^2)-12
  Y_D^2Y_U^2\right] \label{rgel3}\\
\frac{d\lambda_4}{dt} &=& \frac{1}{16\pi^2} \left[
 2 (\lambda_1+\lambda_2)\lambda_4  +
  4(2\lambda_3 +\lambda_4)\lambda_4
  -3\lambda_4(3g^2+g'^{2})
+  3g^2g'^2  \right.\nonumber\\
&& \left.  +6\lambda_4(Y_U^2+Y_D^2)+12
  Y_D^2Y_U^2\right] \label{rgel4}~.
\eear
The solution to the RGEs would respect the approximate relations
\be
\lambda_1 \simeq\lambda_2\simeq\lambda_3\simeq-\lambda_4~, 
\label{flooprel}
\ee
These would be obtained if one naively computes the $\lambda_i$'s using only the fermion loops induced by 
the Yukawa interactions (\ref{lboson1}).
Although the RGEs receive additional contributions to the evolution, such as fermion loops renormalizing the 
scalar wave-functions, these do not greatly modify these relations.
Incidentally, the relations (\ref{flooprel}) among the scalar self-couplings satisfy
the stability condition for the potential (\ref{pot1}), which require
that  $\lambda_1,\lambda_2>0$, with
$\sqrt{\lambda_1\lambda_2}>-\lambda_3-\lambda_4$. 
 The solution  for the $\lambda_1$ is shown
in Figure~\ref{lambda1}, for various values of the cutoff mass scale $M_G$. 

With the above solutions for the Yukawa and scalar self-couplings, we
can readily compute the scalar  mass spectrum. 
Defining the ratio of VEVs as 
\be
\tan\beta \equiv\frac{v_D}{v_U}~,
\label{tanbeta}
\ee
we can express the scalar eigenvalues  and eigenstates in terms of
$\beta$ and $v= \sqrt{v_U^2 + v_D^2}$. 

In terms of the charged components of the $\Phi_U$ and $\Phi_D$
original doublets, the mass-eigenstate charged scalars are 
\be
H^\pm = \Phi_D^\pm\cos\beta - \Phi_U^\pm\sin\beta\label{charged}~,
\ee
with masses given by
\be
M^2_{H^\pm} = -\frac{2\mu_{UD}^2}{\sin 2\beta} - \frac{\lambda_4}{2} v^2 \label{mch}~.
\ee
The neutral states are 
\bear
H&=&
\sqrt{2}\left(Re[\Phi_U^0]\cos\gamma+Re[\Phi_D^0]\sin\gamma\right] \label{h1}\\
h&=&
\sqrt{2}\left(-Re[\Phi_U^0]\sin\gamma+Re[\Phi_D^0]\cos\gamma\right] \label{h2}~,
\eear
with masses 
\bear
M^2_{H} &=& \left(\lambda_1 v^2\cos^2\beta
  -\mu_{UD}^2\tan\beta\right)\cos^2\gamma + 
\left(\lambda_2
  v^2\sin^2\beta-\mu_{UD}^2\cot\beta\right)\sin^2\gamma\nonumber\\
& & +\left[\mu_{UD}^2 + \frac{1}{2}(\lambda_3+\lambda_4)v^2\sin
  2\beta\right]\sin 2\gamma\label{mh1}\\
M^2_{h} &=& \left(\lambda_1 v^2\cos^2\beta
  -\mu_{UD}^2\tan\beta\right)\sin^2\gamma + 
\left(\lambda_2
  v^2\sin^2\beta-\mu_{UD}^2\cot\beta\right)\cos^2\gamma\nonumber\\
& & -\left[\mu_{UD}^2 + \frac{1}{2}(\lambda_3+\lambda_4)v^2\sin
  2\beta\right]\sin 2\gamma\label{mh2}~.
\eear
The pseudo-scalar mass eigenstate defined by
\be
A= \sqrt{2}\left(Im[\Phi_D^0]\cos\beta -Im[\Phi_U^0]\sin\beta\right)\label{afield}~,
\ee
 has a  mass is given by
\be
M_A^2 = -2\frac{\mu_{UD} ^2}{\sin 2\beta}\label{ma}~.
\ee
The mixing angle $\gamma$ is determined by the mass mixing induced by
instantons, $\mu_{UD}$, as well as from the terms in the potential
involving $\lambda_3$ and $\lambda_4$ which result in $\Phi_U-\Phi_D$
mixing when two of the fields are replaced by their VEVs. Is given by
\be
\tan 2\gamma = \frac{\mu_{UD}^2 + (\lambda_3+\lambda_4)v^2\sin
  2\beta/2}{\mu_{UD}^2 +\lambda_4v^2\cos 2\beta/2}\label{tan2gamma}~.
\ee
Finally, the renormalized up-down mixing mass parameter is given in
terms of the instanton parameters as well as the renormalized
self-couplings by
\be
\mu_{UD}^2 = \frac{k\,v^2}{2 M_G^2}
  \frac{\lambda_1\lambda_2\cos^2\beta\sin^2\beta}
{\left[1-kv^2(\lambda_1\cos^2\beta\cot\beta+\lambda_2\sin^2\beta\tan\beta)/(2 M_G^2)\right]}
\label{mud2}~.
\ee

We can now compute the scalar spectrum by defining a value for
$\tan\beta$, the mass of the color octet $M_G$, and the order one
parameter $k$ from the instanton effects. In the context of the model
presented here, it is clear that we will have
\be
\tan\beta\simeq 1\label{tanb1}~,
\ee
given that we assume that the new strong interaction does not violate
isospin. Furthermore, choosing $M_G$ between $(2-4)$~TeV is reasonable
to maintain naturalness in the dynamical mechanism to generate the
electroweak scale. Finally, the instanton parameter is typically of
order one: $k\simeq (0.1-1)$.

 Within this range of parameters, the
scalar spectrum varies little: typically we obtain scalar masses
$M_{H_{1,2}}\simeq (500-600)$~GeV, and similarly for the charged
scalar masses $M_{H^\pm}$.  On the other hand, the pseudo-scalar mass $M_A$ is much
lighter in the range $M_A\simeq (20-200)$~GeV, and crucially depends on the instanton parameter $k$, as well as the suppression with the 
cutoff mass scale $M_G$. 
These results are
summarized in Table~1. 
\begin{table}
\begin{center}
\begin{tabular}{|c|c|c|c|}
\hline
  & $M_G=2~$TeV &$M_G=3~$TeV &$M_G=4~$TeV \\
\hline
$M_A$ &(26-118) &(15-59) & (10-39)\\
\hline
$M_{h}$ &((548-580) &(459-467) & (422-425)\\
\hline
$M_{H}$ &(651-732) &(530-537) &(482-585)\\
\hline
$M_{H^\pm}$ & (603-719)& (495-512) & (453-459)\\
\hline
\end{tabular}
\caption{The scalar spectrum for various values of the cutoff scale $M_G$. The ranges correspond to the instanton parameter range $k=(0.1-1)$ and 
are given in GeV.}
\end{center}
\label{table1}
\end{table}
This is to be  expected, since the scalar
behaves as a pseudo-Nambu-Goldstone boson of the Peccei-Quinn
symmetry, and its mass is only lifted by the explicit breaking due to  instanton effects,
which are dominated by the new interaction at the TeV scale. Thus, we see that
the spectrum of the two Higgs doublet sector we obtain dynamically here
is quite distinct and has all scalars in the $500~$GeV range, with
pseudo-scalar $A$ being much lighter, perhaps as light as direct
bounds allow. This inverted spectrum has important consequences for
the phenomenology of the scalar sector at colliders, as well as on the
bounds, both direct and indirect from precision measurements. 
We comment on both of these in what follows.

\vskip0.5cm
\section{Electroweak Precision Constraints}
\label{ewpc}
The presence of a new fermions results in additional contributions 
to electroweak precision observables. For illustrative purposes, we consider the case of a chiral
fourth generation since it will involve also leptons, and it has a fixed
number of degrees of freedom. The more general case is not necessarily
as constraining on the parameter space.
 In addition to the fermionic contributions to precision observables,
 the formation of a composite two-Higgs doublet sector at low energies
 adds new contributions. The contributions to the $S$ and $T$
 parameters from the two-Higgs doublet sector can be found, for
 instance, in Refs.~\cite{thdmgen} and ~\cite{st2hdm}. 
The typical spectrum obtained by the double fourth-generation condensation, with a light pseudo-scalar and heavy and almost degenerate charged and CP-even scalars, results in a modest positive contribution to $S$. This is typically of the order of $S_{2H}\simeq 0.1$.
On the other hand, the contributions of the scalar spectrum to $T$ can vary more, approximately in the range 
$-0.2 <T_{2H}<0.2$, for values of the cutoff in the range $\Lambda=(2-4)~$TeV , the instanton 
parameter taken in the interval $k=(0.1-1)$, and $\tan\beta=0.9$.

These contributions must be added to the ones coming from the
fourth-generation fermion loops~\cite{hesu,kribs}.
 Since the composite scalar sector is made out of fourth-generation quarks, in principle 
these contributions to $S$ and $T$ are not completely
independent. However, if we think in terms of a large-$N_c$ expansion,
the scalar contributions should be suppressed with respect to the
single-quark loops. We will then add the two contributions
coherently. 
The addition of these two contributions to $S$ and $T$ was done for
much more  general scalar and fermion spectra in Ref.~\cite{hesu}. 
In Figure~2 we plot the parameter space of the model considering the cutoff 
region $\Lambda=(2-4)~$TeV, the instanton parameter varying in the range $k=(0.1-1)$, 
and $\tan\beta=0.9$. The RGE results from the previous section determine the overall scale of the fourth-generation quark spectrum. We take then the isospin splitting, which is mostly induced by electroweak corrections, to  vary between $(0-100)~$GeV.  
 In general, the overall mass scale of the lepton sector is not set by $\Lambda$, although it is related to it in some specific models~\cite{bd1}.  To be general, we take neutrino masses to vary in the range $m_{\nu_4}=(100-500)~GeV$, and again the isospin splitting with the charged lepton to be in the interval $(0-100)˜$GeV.
The ellipses represent the $68\%$ and $95\%$ C.L. bounds from experiment as obtained in 
Ref.~\cite{gfitter}. 
\begin{figure}
\begin{center}
\includegraphics[scale=1.1]{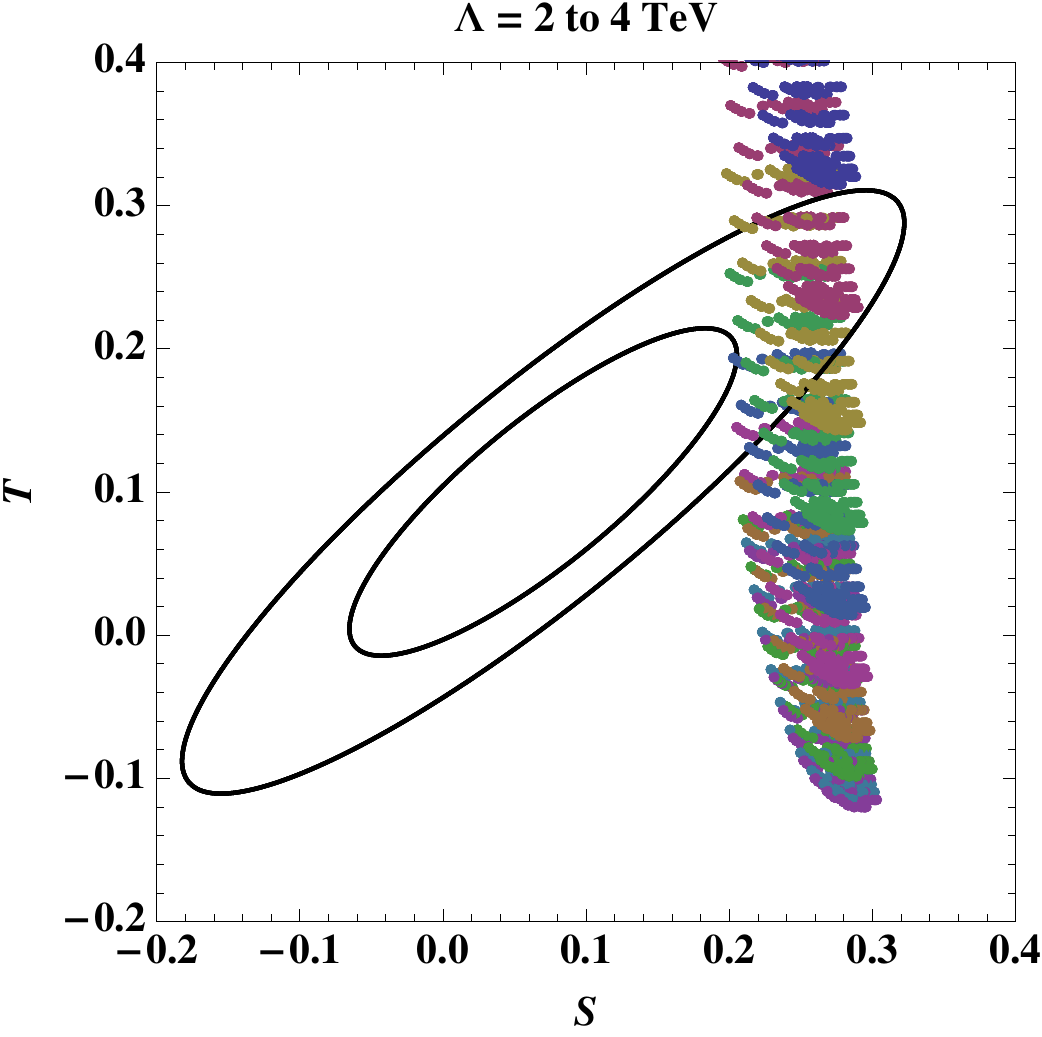}
\caption{The contributions to the S and T parameters from a chiral
  fourth-generation and the composite two-Higgs sector 
resulting form quark condensation.} 
\end{center}
\label{stplot:fig}
\end{figure}
We can see that there is a region of parameter space of the model that is within the $95\%$ C.L.
interval. Thus, the scalar spectrum presented in the previous section, and the fourth-generation  spectrum resulting from fourth-generation quark condensation are not  excluded by electroweak data.  
The more general case, where the new fermions do not carry color,
should be very similar  and we take the case studied above as a
concrete indication of the existence of a sizable  paramater space. 

\section{Phenomenology}
\label{pheno}
In this section we consider some phenomenological aspects of these generic models. 
First, we address the issue of fermion masses and flavor violation. 
We then briefly discuss current bounds on and future searches for the scalar spectrum presented here. 

\subsection{Flavor Conservation and Constraints}
Up to now, the model presented here breaks the electroweak symmetry
and gives masses to the condensing fermions, e.g. the fourth-generation quarks. In order to give masses to all other fermions additional physics must be introduced. In general,  the interaction giving rise to EWSB is not enough to 
generate the SM flavor structure. Higher dimensional operators such as 
\be
\bar f^i_Lf^j_R\bar f^k_R f^\ell_L ~,
\label{hdops}
\ee
are needed. In (\ref{hdops}),  $f^i$ denotes a fermion of generation $i$ in a generic basis not necessarily aligned with the basis of the gauge interactions responsible for fourth-generation
condensation. The size of the coefficients of these operators reflect a new dynamics, possibly at a higher energy scale.  For instance, in the spirit of the model of Ref.~\cite{bd1}, there are four-quark operators generated by the exchange of the KK gluon. These become supercritically strong only for quark zero-modes highly localized in the IR. This is only the case for the fourth-generation. On the other hand, the flavor-violating operators in (\ref{hdops}) are generated 
by non-renormalizable interactions in the 5D bulk, presumably
resulting from a broken gauge flavor interaction at the UV scale. 
More generally, whatever the origin of the higher dimensional
operators necessary to communicate  EWSB to fermion masses, this can
always be accommodated by having the doublet $\tilde\Phi_U$ couple
only to up-type quarks, while having $\Phi_D$ couple only to down-type 
quarks and leptons. This is a two-Higgs model of type-II
~\cite{thdmgen}, and it does not result in tree-level neutral flavor violation.

In principle, we could worry about the fact that $\Phi_U$ and $\Phi_D$ are not mass eigen-states, since their mixing $\mu_{UD}$ is generated at low energies by instanton effects. 
However, as already pointed out in Ref.~\cite{fgcon}, the running of the dimensionless Yukawa couplings is unaffected by the mixing term, which only represents a soft breaking of the PQ symmetry. As a consequence, running to low energies preserves the fact that $\Phi_U$ and 
$\Phi_D$ only couple to up-type and down-type fermions respectively, ensuring that 
after mass diagonalization the neutral scalars have flavor-preserving neutral interactions.

Going beyond tree-level, the physical spectrum of the two-Higgs doublet model 
type II generates new contributions to various flavor observables~\cite{thdmflavor}. 
The most stringent constraint comes from the contribution of the charged Higgs states to 
flavor-changing loop-induced decays, most notably $b\to s\gamma$. The data from B physics
is consistent with the two-Higgs model spectrum as long as  the $95\%$ C.L. bound f
 $M_{H^\pm}> 316~$GeV is satisfied.
In the present models, these states have masses well in excess of this bound, starting at about
$500~$GeV. On the other hand, the lighter pseudo-scalar state does not result in important contributions to loop-induced decays.
We conclude that the two-Higgs doublet model spectrum favored by the
condensation new chiral fermions described in Section~\ref{model} is not 
highly constrained by flavor data at the moment.

\subsection{Direct Searches}
The low energy bosonic spectrum of these models, with very heavy  scalars $(h,H,H^\pm)$ and a rather light pesudo-scalar  $A$, 
presents a very distinct phenomenology. Although spectra of THDMs with $A$ the lightest state  have been studied in the context 
of  supersymmetric models~\cite{nmssm}, as well as more generic THDM studies~\cite{flipped}, the fact that the scalars are typically above $450$~GeV presents 
important differences both for the existing bounds as well as the search strategies at the LHC. Although a detailed study of the phenomenology 
will be carried out elsewhere~\cite{inprepa}, there are some general remarks we can make here. 
\begin{figure}
\begin{center}
\includegraphics[scale=0.8]{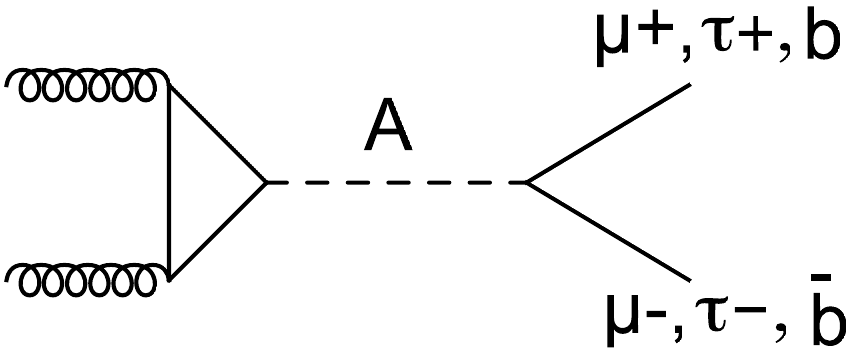}\\~\\
\includegraphics[scale=0.7]{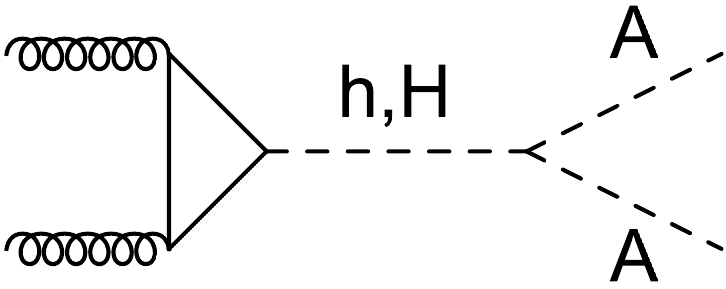}\qquad
\includegraphics[scale=0.7]{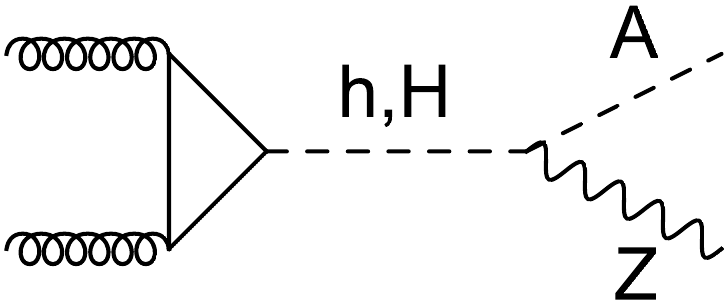}
\caption{Diagrams dominating the production of the pseudo-scalar  $A$ at hadron colliders.} 
\end{center}
\label{aproduction}
\end{figure}
First, existing bounds on $M_A$ rely almost exclusively on its production in association with $h$ at LEP. This bounds do not apply to our spectrum since
$M_h>400~$GeV means that LEP had no kinematic reach.  
In the case the new fermions are fourth-generation quarks, the bounds
on the direct s-channel production of the scalars and pseudo-scalar
are affected by the enhancement of the production cross sections
due to the presence of the fourth-generation in the loop. This gives a
cross section which is typically a factor of $(7-9)$ larger than that
of a standard THDM  type II with  three generations. This is not the
case for the colorless fermion models, where only the
$\phi\gamma\gamma$ couplings are affected significantly.
Considering the light end of the pseudo-scalar state, the bounds from
B factories are $M_A>9.5~$GeV at $90\%$~C.L.
Tevatron searches are not sensitive to this flipped spectrum. The LHC
searches are beginning to be constraining in the presence of a fourth
generation.  However, and although the SM Higgs is excluded up to
masses of $\simeq 600~$GeV in this case, this is yet not applicable to
the two-Higgs doublet spectrum presented here. 
 
As we can see from Table~\ref{table1}, $M_A$ can be anywhere between $10$ and $120$~GeV, and possibly somewhat beyond both ends of this range.  
The s-channel $A$ production gives the largest cross section, but it
represents a challenge, especially at the larger masses where the
branching ratio into b quarks dominates. Although the modes mediated
by the heavy CP-even neutral scalars is suppressed due to their large
masses,  the enhancement of the cross section due to the additional
fermions in the loop makes this modes of great
interest~\cite{inprepa}, especially for models with additional colored doublets.  
The CP-even scalars $(h,H)$, would be produced with a larger cross
section too, but the  addition of new important decay channels such as
$AA$ and $AZ$ change the direct limits obtained for the SM Higgs  decaying
into  $W^+W^-$, $ZZ$ by ATLAS and CMS~\cite{hsearch11}. A full phenomenological study
of the branching ratios and preferred decays for a given spectrum for
both the colored and colorless fermion cases is
left for a separate publication~\cite{inprepa}.

\section{Conclusions and Outlook}
\label{conc}
We have shown that electroweak symmetry breaking by fermion
condensation typically leads to a two-Higgs doublet model at low
energies.  We modeled the scalar spectrum of these scenarios by using
the NJL approach. We obtained an inverted scalar spectrum, with 
the neutral CP-even and charged scalars heavy $(h,H,H^\pm)$, typically at
$(500-600)~$GeV, whereas the pseudo-scalar $A$ is much lighter due to
a Peccei-Quinn symmetry in the original action, only broken by the
instantons of the new strong interactions. 

The low energy scalar spectrum corresponds to a type-II two-Higgs
doublet model. Thus tree-level flavor conservation is built in. 
For the masses obtained, loop-induced flavor-changing processes are
not  binding. mostly since $m_{h^\pm}> 450~$GeV avoids $b\to s\gamma$
bounds.  The study of electroweak precision observables shows that per
se the scalar spectrum does not violate $S$ and $T$ bounds
significantly.  However, when a fourth generation is included
important regions of parameter space are excluded. This exclusion is
independent of whether the new fermion sector is a fourth generation
or  a more exotic matter content, since the contributions to $S$ and
$T$ depend only on the number of fermions, and their mass splittings. 
However, we conclude that an important region of the parameter space is still
allowed by electroweak precision bounds, as can be seen from
Figure~2. 

This inverted two-Higgs doublet model spectrum leads to a distinct
phenomenology at the LHC, with the typical decay channel for the
neutral CP even states, $(h,H)\to WW, ZZ$ now being in competition
with $(h,H)\to AA, AZ$. These new channels, involving either 4 b's,
$\tau$'s or 2 b's and 2 charged leptons, should be looked  for at the
LHC, for the
appropriate regions of the model's parameter space. We expect that
the LHC with $\sqrt{s} =7$~TeV  will be able to fully test these
regions.   
A detailed study of this phenomenology will be 
undertaken in Ref.~\cite{inprepa}.

\bigskip

{\bf Acknowledgments:}
The authors acknowledge the support of the State of S\~{a}o Paulo
Research Foundation (FAPESP), and the Brazilian  National Counsel
for Technological and Scientific Development (CNPq).


\newpage

\end{document}